\documentclass[conference]{IEEEtran}
\usepackage{amsmath}
\usepackage[utf8]{inputenc}
\usepackage[english]{babel}
\usepackage{hyperref}
\usepackage{siunitx}
\usepackage{graphicx}
\usepackage{float}
\usepackage{threeparttable}

\usepackage[
backend=biber,
style=numeric,
sorting=nty
]{biblatex}

\IEEEoverridecommandlockouts

\begin{document}

\date{}

\title{Attacking Hardware AES with DFA}

\author{
\IEEEauthorblockN{Yifan Lu\thanks{This work was not supported and do not represent the approval or rights of any third parties.}}
\IEEEauthorblockA{\href{mailto:me@yifanlu.com}{me@yifanlu.com}}
} 

\maketitle

\begin{abstract}
We present the first practical attack on a hardware AES accelerator with 256 bit embedded keys using DFA. We identify the challenges of adapting well-known theoretical AES DFA models to hardware under attack from voltage fault injection and present solutions to those challenges. As a result, we managed to recover 278 real-world AES-256 keys from a secure computing system in a matter of hours with minimal cost.
\end{abstract}

\section{Introduction}

Although there is a wealth of work in differential fault analysis (DFA) attacks on AES~\cite{Breier14} and it is well understood that such attacks works on hardware AES accelerators~\cite{Dusart02}, there has been few practical attacks on real-world targets. In 2012, Sony released their second hand-held gaming console, the PlayStation Vita. Although it was not the runaway success of its predecessor~\cite{USGamer}, Sony thoroughly improved the software and hardware security features on their new console~\cite{Lu13}. At the root of their secure boot system is a cryptographic accelerator (only accessible by a dedicated security CPU) that operates with keys embedded in the silicon which are not directly accessible by software. The keys can only be referenced through hardware protected key-slots. By obfuscating the keys this way, the designers hope that in the event that the system is compromised and attackers wish to use the device as a black-box to decrypt data, they can reference new key-slots in a firmware upgrade and re-secure the system. This works as long as both of the following points hold true:

\begin{enumerate}
    \item \textit{The system is not compromised before it locks out the important key-slots.} The Vita will always revoke the permission to use a key-slot when it is no longer needed. Unused key-slots are also locked down early in the boot process.
    \item \textit{The keys themselves cannot be extracted by the attacker even if she compromises the secure processor.} Otherwise, the permissions enforced by the cryptographic accelerator can be bypassed.
\end{enumerate}

It has already been demonstrated~\cite{Lu19} that assumption 1 is broken with fault injection attacks on the secure processor (called ``F00D'') and therefore the security has already been defeated. However, we wish to go farther and break assumption 2 as well. We do just that with a DFA attack on the Vita's cryptographic accelerator (which we nicknamed ``Bigmac'').

\subsection{AES}

The Rijndael cipher~\cite{aespaper}, known more commonly as AES, is a substitution permutation based cipher that is widely used for encryption to ensure confidentiality of data. The cipher operates in $N$ rounds where $N$ is $10$ for AES-128 and $14$ for AES-256. In each round except the last, there are four operations performed on a $4 \times 4$ state matrix which is initialized before the first round by each plain-text byte XORed with the first round key. There are $N+1$ round keys generated through a separate process not described here. The state operations are defined briefly:

\begin{enumerate}
\item \texttt{SubBytes} is the substitution step where a non-linear function is applied on the input byte.
\item \texttt{ShiftRows} performs a cyclic rotation on each row of the state.
\item \texttt{MixColumns} linearly combines the elements in each column. It can be represented as a multiplication of each column with a constant matrix. This step is skipped for the last round.
\item \texttt{AddRoundKey} ties the result to the key by XORing each element with an element from the current round key.
\end{enumerate}

\subsection{DFA}

It would be remiss to not start with a reference to Boneh, Demillo, and Lipton's 1997 paper~\cite{Boneh97}. The authors described how an incorrect RSA signature produced by faulty hardware can be used to retrieve the private key. Shortly thereafter, Biham and Shamir~\cite{Biham97} discovered that faulty results from symmetric encryption systems like DES can also be used to extract the secret key. They called the attack Differential Fault Analysis because they used the information gleaned from related ciphertexts produced from good hardware and faulty hardware to find the secret key.

DFA can also be applied to AES~\cite{Dusart02}. According to Dusart et al., if the fault is modeled by a single unknown byte, $\epsilon$, which is XORed into a specific element of the state matrix before \texttt{MixColumns} of round $N-1$, then one can solve for four bytes of the round $N$ key. The high level idea is that the non-linear structure of the S-Box can be abused to leak information about the state. As an example, Dusart presented the following system of equations for a fault $\epsilon$ at a fixed location in round $N-1$:

\begin{equation}
\left \{ \begin{aligned}
s(x_0 + 2 \epsilon) &= s(x_0) + \epsilon'_0 \\
s(x_1 +   \epsilon) &= s(x_1) + \epsilon'_1 \\
s(x_2 +   \epsilon) &= s(x_2) + \epsilon'_2 \\
s(x_3 + 3 \epsilon) &= s(x_3) + \epsilon'_3 \\
\end{aligned}
\right .
\end{equation}

The S-Box of \texttt{SubBytes} is represented as $s(x)$ and the unknowns are $x_0, x_1, x_2, x_3, \epsilon$. The observed faulty ciphertext difference are $\epsilon'_0, \epsilon'_1, \epsilon'_2, \epsilon'_3$. With each fault, we observe a different set of $\epsilon'$ and assume a different unknown $\epsilon$ but $x$ does not change. After enough samples, we reduce the solution set to $1$ and solve for $x$. Once the state bytes are revealed, it is easy to extract four bytes of the round $N$ key. Repeat the whole procedure with faults at different offsets, and it is possible to recover the entire round $N$ key and going from the round $N$ key to the original key is just a matter of reversing the key scheduling algorithm (which is not secret). Dusart et al. were able to extract an AES-128 key by "analyzing less than 50 ciphertexts."

Recent progress in AES DFA theory since then has been in reducing the number of needed ciphertexts~\cite{Ali12, Kim12, Tunstall} as well as relaxing the fault model to handle more faults (or earlier faults)~\cite{Kim12, Moradi06, Piret03}. There has also been attacks targeting the key schedule rather than the round states~\cite{Floissac11}. In practical attacks, AES DFA has been shown to work on ARM processors~\cite{Barenghi10}, on FPGA~\cite{Ali10}, and on an ASIC~\cite{Selmane08}.

\subsection{Prior Work}

Our work is closest in design to Selmane's~\cite{Selmane08} for a smart-card. Their target was also a dedicated AES accelerator running alongside a CPU. They also used Piret's model~\cite{Piret03} for performing the DFA attack. The method for inducing the faults was timing violations. However, one key difference is that they created the timing violations by under-powering the smart-card such that the AES operations will be faulty but the CPU can still operate correctly. Our method for creating timing violations is from voltage glitching~\cite{Lu19}. Under-powering does not work for us because the global critical paths are not inside Bigmac. When we try to under-power the device, the F00D security processor will not execute code, and we cannot control Bigmac. Additionally, by using a voltage fault injection, we can create precise glitches that target specific AES rounds. This is important for reducing the number of needed faulty ciphertexts.

We used Piret's model\footnote{It should be noted that more than a decade of work has taken place since Piret's results and there exists many AES DFA models that have lesser requirements on the required faults. However, we were unable to find any open-source implementation of the newer ideas. Because we can control precisely where the fault takes place, we can make use of this constrained model and it saves us the time of implementing our own DFA tool from scratch.} of a single fault between the \texttt{MixColumns} of round $N-3$ and $N-2$ (which itself is an extension of the original model from Dusart described above). This model only requires two faulty ciphertexts and the correct ciphertext to recover the final round key. We used an implementation of the attack on this model called phoenixAES~\cite{phoenixAES}. This tool was developed as a way of attacking white-box AES~\cite{Bos15}, but we find that it can also process hardware generated faults without any modification.

Because it is not possible to characterize each faulty ciphertext as ``good'' or not (i.e. fits the assumptions of the model) without knowing the key, we developed a brute force approach to try every pairing of ciphertexts. We believe this is similar to an approach taken by Riscure~\cite{Riscure} but was unable to verify this because they did not go into details about their implementation.

\section{Fault Injections}

Piret's fault model requires exactly one byte of the AES state to be corrupted between the \texttt{MixColumns} operation of round $N-3$ and $N-2$~\cite{Piret03}. To recover the round $N$ key, we need two different single-byte corruptions of this type. The first practical challenge is to create fault injections that can meet this requirement. We used voltage glitching to inject the fault because of previous success~\cite{Lu19} in voltage glitching the F00D processor on the same target. Specifically, we applied the crowbar voltage glitching technique~\cite{OFlynn16} because of its low cost and high applicability.

\subsection{Hardware AES}

Although we do not know the exact design of Bigmac's AES implementation, we can reasonably assume that it is optimized in some way. We know that \texttt{SubBytes} and \texttt{MixColumns} (as well as their inverse) are the most expensive operation~\cite{Nalini09} and therefore it is highly likely that the critical paths are within those operations. This means that with the right timing, we can achieve a glitch that only affects one operation and that it is possible to meet the requirement for a one byte corruption.\footnote{Ideally we can glitch only for the duration of one operation, but optimized AES implementations typically perform all four operations of a round in one or two cycles. However, because we know that the path length for each operation is not equal, we do not run into the awkward situation of faults occurring in multiple operations, which does not meet our requirement for a single byte fault.}

From a software perspective, Bigmac has a simple command interface. It has memory-mapped registers accessible only by the F00D security processor at physical address \texttt{0xE0050000}. The ARM based application processor cannot see this address range at all and must interface with F00D to use Bigmac indirectly. Only TrustZone on the ARM processor can communicate with F00D. Bigmac has support for AES block modes CBC, ECB, and CTR, with key-sizes of 128, 192, and 256. In addition, it also supports AES-CMAC, HMAC-SHA, SHA, memcpy, memset, and generating random numbers. To perform a Bigmac operation, F00D passes in a source and destination address, the length of the data, the operation, and (when required) the address of the IV. For keyed operations, Bigmac accepts either a fixed key written to a set of input registers directly or the index of a key-slot. It is also possible to set the output destination to a key-slot instead of a memory address in some cases.

Each key-slot has a permission bit associated with it. Some key-slots are only allowed to encrypt data to another key-slot. We call these ``master'' key-slots. It is normally not possible to directly observe ciphertext produced by encrypting with a master key. There are $30$ master keys in the PlayStation Vita, some of which are device unique and others which are common to all Vita devices. There are $250$ additional non-master key-slots (some device unique) not derived by software that we can directly observe the ciphertext. Finally, the remaining key-slots are either derived from the master keys along with data from the firmware or loaded directly from software decrypted by Bigmac. Key-slots can also be disabled such that they cannot be used until the next reset. Most keys including all master keys are disabled early in boot before the operating system is loaded.

The results of this paper include the procedure we devised to obtain $248$ non-master keys and all $30$ master keys by leveraging DFA, code execution through voltage glitching, a hardware vulnerability, and the computation power of about 500 core-hours.

\subsection{Glitch Parameters}

The effects of crowbar voltage glitching depends largely on two factors: \textit{when} the crowbar circuit is activated and \textit{how long} it stays activated~\cite{OFlynn16}. To reduce the variance in our measurements, we replace the target's external clock input with our own clock running at $f = \SI{12}{\mega\hertz}$. Bigmac runs with a clock derived from the external clock with frequency $f_c$. From measurement of Bigmac's AES timing, if we assume that one AES round takes one clock cycle, then $f_c = f$. Our glitching hardware also runs with a clock derived from the same source with frequency $f_g = 4 f$. When we refer to ``cycles'', it is in units of $1/f$.

Since we are dealing with AES-256, it is necessary to retrieve two round keys in order to recover the full key~\cite{Dusart02}. This means we need to find two sets of parameters: $n_{N-2}, m_{N-2}$ and $n_{N-3}, m_{N-3}$. We define $n$ to be the offset from a fixed trigger signal before the AES engine starts to the crowbar activation and $m$ to be the duration of the crowbar activation. Since we are applying Piret's model, this means we have to target round $N-2$ and round $N-3$.

\section{Setup}

Boot time code execution is a prerequisite for interfacing with Bigmac before the target key-slots are disabled. We reproduce the setup described in~\cite{Lu19} to achieve this. As such, we make use of the ChipWhisperer Lite, an open source hardware fault injection and side channel analysis tool. We designate, through a series of scripts, two separate modes of operation. In boot mode, we configure the ChipWhisperer to perform the previously described voltage glitch attack on F00D to gain early boot execution. Once that succeeds, we load an RPC payload that interfaces with the ChipWhisperer through UART serial and enter DFA mode. In DFA mode, we send the plaintext through the serial port and use the RPC interface to set up Bigmac and toggle a GPIO pin before starting the AES operation. Then we perform a voltage glitch using ChipWhisperer by waiting $n$ cycles after a GPIO toggle to activate the glitch circuit and turn it off after $m$ cycles. The device will then return the output ciphertext through the serial port. The same glitching hardware is used with different configurations for the two modes. We will only describe our setup for DFA mode.

\subsection{Reducing Capacitance}

To minimize the impact of the power distribution network (PDN), we make a couple of modifications to the PCB. First, we trace and remove every decoupling capacitor to the core 1.1V voltage domain. Figure~\ref{fig:decoups} shows the capacitors that were removed.

Next, we introduce a $\SI{10}{\ohm}$ shunt resistor\footnote{Originally we attempted to do a DPA attack but gave up after a couple of months without any results. We believe that because the AES engine was designed for power efficiency, the SNR was too low to get accurate measurements for this device and $\SI{10}{\ohm}$ was the highest shunt we can choose that still allows the device to operate. The same shunt resistor was used for the DFA attack because it was already in place. There is no solid evidence that the shunt resistor is needed for DFA, but empirical results show that without the shunt resistor in place, the minimum width of the glitch needs to be about 3x as much for the system to show any faulty behavior.} by cutting the trace from the device's own power management chip to the main SoC (figure \ref{fig:cuttrace}). We designed a simple board (figure~\ref{fig:psvcw}) that contains the shunt resistor, a filter capacitor, and ports for an external power supply, a measurement probe, and a SMA connector to the ChipWhisperer to perform the voltage glitch.

\subsection{Measurements}

\begin{figure}[t]
\centering
\includegraphics[width=8cm]{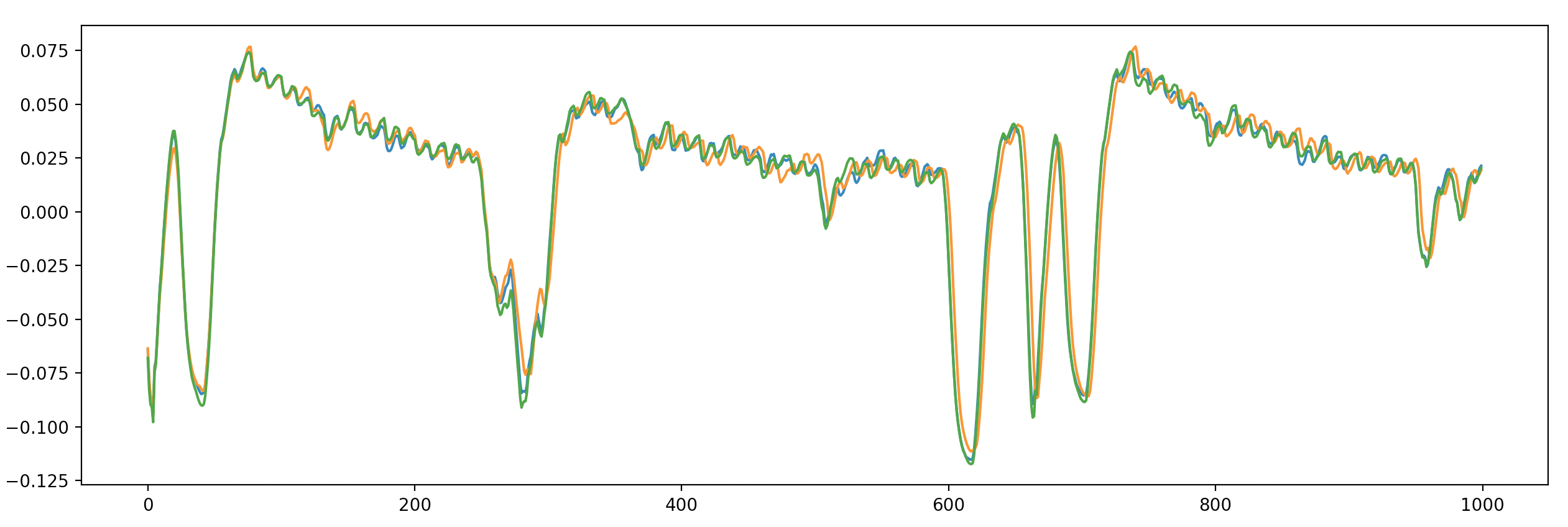}
\caption{1000 samples at \SI{12}{\mega\hertz} done three times. The GPIO toggles on at 0 and off at 600. The entire 14 rounds of AES takes place in the ``dip'' around 250.}
\label{fig:aesops}
\end{figure}

On the target device, we use our RPC interface to toggle a GPIO pin and immediately start the Bigmac AES operation. Using a CW501 differential probe for ChipWhisperer sampling at $f$, we capture the power trace triggered by the GPIO pin. Using these traces, we can get a rough idea of when the AES operation takes place.\footnote{Another reason for the low SNR that made DPA difficult was due to the fact that the DMA reads and writes dominates the spikes we see (not AES rounds). We confirmed this by running AES-ECB and AES-CBC and noticed that two extra spikes appear for reading and writing the IV from memory. After setting the destination to a key-slot (instead of SRAM), we observe half as many spikes.} Figure~\ref{fig:aesops} shows what the traces look like.

\subsection{Encrypt vs Decrypt}

When we targeted the AES-ECB-256 encrypt operation, we ran into issues with the corrupted ciphertext that we suspect was due to both the AES state operation and the key scheduling being faulted at the same time.\footnote{Our simple analysis fails if the key schedule was also faulty.} It is likely that the key scheduling was taking place in parallel with the encrypt operation. When we target the AES-ECB-256 decrypt operation, the issues we observed dramatically decreased, which confirms our hypothesis. As a matter of convention, we will continue to refer to the AES engine output as the ``ciphertext'' and the round number in reference to the encryption round even though our fault injection was on the decrypt operation.

\subsection{Simple Analysis}

By injecting faults at offsets $240 < n < 280$ and width $m=1$ we were able to observe faulty ciphertexts. However, most of the faulty texts were not at the right round and therefore cannot be used for Piret's DFA model. Normally, this would not be an issue as we can just throw out results that fail the attack, however for reasons that will be made clear later, it is to our advantage that we maximize the probability that each faulty ciphertext could be ``useful.'' To do this, we need to constrain $n$ more and identify the precise round that each $n$ value affects. Following the idea in~\cite{Zussa13}, we setup Bigmac with a known key and then perform a faulted decrypt operation. Then we try to encrypt (the inverse operation) the faulty ciphertext with the same key and identify the first step where the state got corrupted.

\begin{figure}[t]
\centering
\includegraphics[width=8cm]{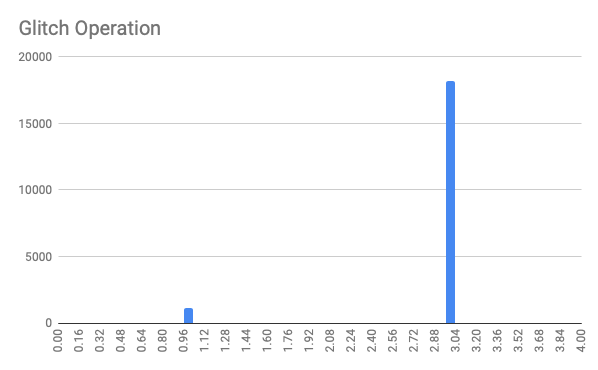}
\caption{Distribution of the round operation faulted by the glitch. The operation number is in order of an encrypt round. $1$ is \texttt{SubBytes} and $3$ is \texttt{MixColumns}.}
\label{fig:glitchop}
\end{figure}

\begin{figure}[t]
\centering
\includegraphics[width=8cm]{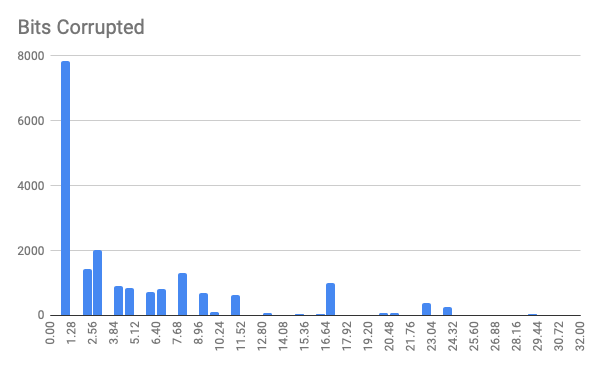}
\caption{Distribution of the number of bits in the state corrupted. The majority of the corruptions are a single bit.}
\label{fig:bitscorrupted}
\end{figure}

More specifically, we design an analysis script that performs two AES encrypt operations in parallel: one on the expected ciphertext and one on the faulty ciphertext. After each step, we count the number of bits that differ in the state, and we return the round and step that has the least number of different bits. This works with high probability because AES is designed to have diffusion, so each step after the fault would, on average, be more different from the step before it. Table~\ref{tab:analysis} shows some example output from this analysis. Figure~\ref{fig:glitchop} shows the distribution of where the fault was seen and figure~\ref{fig:bitscorrupted} shows the distribution of the number of bits corrupted.

\begin{table}[]
\centering
\caption{Sample Analysis}
\label{tab:analysis}
\begin{tabular}{llllll}
\textbf{Decrypt Output} & $\boldsymbol{m}$ & $\boldsymbol{n}$ \\
\textbf{Corrupted Bit Mask} & \textbf{Round} & \textbf{Operation} \\
\hline
\scriptsize\texttt{9E8EDBEBE1CF276208912BB325CF6E7F} & - & - \\
\scriptsize\texttt{00000000000000000000000000000000} & - & - \\
\hline
\scriptsize\texttt{5FF5D6AEADFF594817F4FB3F565EB5F1} & 282 & 1 \\
\scriptsize\texttt{00000000000000400000002000000000} & 3 & \small\texttt{MixColumns} \\
\hline
\scriptsize\texttt{6D139A0FB71775A4C55F8E6C2B88162B} & 281.5 & 1 \\
\scriptsize\texttt{00004000000000000000000000000000} & 4 & \small\texttt{MixColumns} \\
\hline
\scriptsize\texttt{F3B414E25E4CF5B1D7CEA101C61A9A3C} & 281.5 & 1 \\
\scriptsize\texttt{00014203000020200000000000000020} & 4 & \small\texttt{MixColumns} \\
\hline
\scriptsize\texttt{9D7CEF8A3B9E222FAC826D6E21BC6BC3} & 279.5 & 1 \\
\scriptsize\texttt{00104000000000000004101200000060} & 7 & \small\texttt{MixColumns} \\
\hline
\scriptsize\texttt{A1A68EFB05B99D0E7C1C18328265F2BD} & 279.5 & 1 \\
\scriptsize\texttt{00104000000000000000000000000020} & 7 & \small\texttt{MixColumns} \\
\hline
\scriptsize\texttt{621A9AE2F689F316DC1C8BA8F5794C4C} & 277.5 & 1 \\
\scriptsize\texttt{00004000000000000000000000000000} & 10 & \small\texttt{MixColumns} \\
\hline
\scriptsize\texttt{53BA4B36688166424E5E7ACEFBDF8357} & 276.75 & 1 \\
\scriptsize\texttt{18021209000000000000000000000020} & 11 & \small\texttt{MixColumns} \\
\hline
\scriptsize\texttt{48E042EB3A7E7015C8293C85089F615E} & 275.5 & 1 \\
\scriptsize\texttt{00000000000000000000000000000004} & 13 & \small\texttt{MixColumns}
\end{tabular}
\begin{tablenotes}
\small
\item Selected sample analysis results from AES-256 decrypting input \texttt{00000000000000000000000000000000} with key-slot \texttt{0x3FF}. The first row is the expected output while the remaining rows are results from a fault at offset $m$ with width $n$. The mask shows which bits in the AES state was corrupted. The round and operation is the step where the AES state was corrupted.
\end{tablenotes}
\end{table}

We notice that for certain value of $m$ (such as the ones shown in table~\ref{tab:analysis}), we are able to cause faults in the \texttt{MixColumns} of a specific round the vast majority of the time. This is evidence that our glitching setup is robust and precise enough to perform Piret's DFA attack.

\begin{table}[]
\centering
\caption{Glitch Parameters}
\label{tab:offsets}
\begin{tabular}{llll}
          & \textbf{AES-128} & \textbf{AES-256} & \textbf{AES-256} \\
          & \textbf{Fixed} & \textbf{Non-master Slot} & \textbf{Master slot} \\
          \hline
$n_{N-2}$ & $270.75$                                                                    & $271.5$                                                                              & $282.25$                                                                         \\
          \hline
$n_{N-3}$ & $270.25$                                                                    & $272.25$                                                                             & $281.5$                                                                         
\end{tabular}
\begin{tablenotes}
\small
\item Offsets found for three kinds of operations: AES-128 encrypt with a fixed key (only used for debugging our setup), AES-256 decrypt with a non-master key-slot, and AES-256 decrypt with a master key-slot. Glitch width is $n = 1$ for all cases.
\end{tablenotes}
\end{table}

\section{Attack}

With the offsets found from the analysis (table \ref{tab:offsets}), we can collect faulty ciphertext which are (with high probability) from the required round. However, there is no guarantee that the faults are only a single byte or that we can cause two different single byte faults per round. Therefore, we need to do some post-processing.

\subsection{Filtering out multi-byte corruptions}

Recall that Piret's model requires two different single byte corruptions to recover a single round key. However, as observed in table~\ref{tab:analysis} and figure~\ref{fig:bitscorrupted}, the structure of our faults appear to be 1-5 bits flipped. If all the bits flipped are inside a single byte, then we are good. However, it is clear that not every faulty ciphertext has the bit flips confined to a single byte.

Fortunately, this is an easy problem to solve. The DFA attack will fail (no solution to the equations) if the wrong faulty ciphertexts are used. Therefore, if we collect $M$ faulty ciphertexts, we can just attempt the DFA attack with all ${M \choose 2} = O(M^2)$ possible pairings.

\subsection{Using multi-byte corruptions}

The above works well, and we were able to recover a many key-slots with enough samples. For the remaining key-slots which we were unable to find a pair of required faulty ciphertexts after hours of sampling. We have to improvise another workaround.\footnote{We could have implemented a different DFA fault model which is less constrained but it was easier to improvise a more inefficient method.}

First, notice from table~\ref{tab:analysis} that the mask of corrupted bits show some bits are more likely to be flipped than others---even if the fault happens in different rounds. The explanation for this phenomenon is the physical hardware data path for each bit of the state is not equal. We mentioned previously that certain operations (such as \texttt{MixColumns}) are more likely to fault because the data paths are longer than those of other operations and that makes a timing violation more likely. However, even within \texttt{MixColumns}, there are differences in the data path for each bit of the state.\footnote{Predicting which bit is more likely to flip is difficult because it is data dependent as well as process dependent.}

Given a collection of faulty ciphertexts, we define a ``static fault'' to be any bit corruption that is common to all the ciphertexts and a ``dynamic fault'' to be the bit corruption(s) that only occur in only some faulty ciphertexts.

\subsubsection{Second Order DFA}

We claim that any number of static faults do not affect the results of DFA. Taking any existing AES DFA technique and it is possible to relax the requirement for a ``correct'' ciphertext to that of one that only contains static faults.

We prove this claim for Dusart's example DFA attack on round $N-1$. Let matrix $S_{r,\text{Op}}$ be the correct state at round $r$ with operation $\text{Op}$ and let $F_{r,\text{Op}}$ be the faulty state matrix. Matrix $Z$ contains the static faults (up to 16 bytes) and $A_0$ is the matrix constant for \texttt{MixColumns}.

\begin{equation}
\begin{aligned}
Z &= \begin{pmatrix}
\zeta_{1} & \zeta_{2} & \zeta_{3} & \zeta_{4} \\
\zeta_{5} & \zeta_{6} & \zeta_{7} & \zeta_{8} \\
\zeta_{9} & \zeta_{10} & \zeta_{11} & \zeta_{12} \\
\zeta_{13} & \zeta_{14} & \zeta_{15} & \zeta_{16} \\
\end{pmatrix} \\
A_0 &= \begin{pmatrix}
2 & 3 & 1 & 1 \\
1 & 2 & 3 & 1 \\
1 & 1 & 2 & 3 \\
3 & 1 & 1 & 2 \\
\end{pmatrix}
\end{aligned}
\end{equation}

Let $\epsilon$ be a single dynamic fault at byte $0$. We show the effect of the faults in the last two rounds:

\begin{align*}
F_{N-1,\text{ShiftRows}} &= S_{N-1,\text{ShiftRows}} + Z + \begin{pmatrix}
\epsilon & 0 & 0 & 0 \\
0 & 0 & 0 & 0 \\
0 & 0 & 0 & 0 \\
0 & 0 & 0 & 0
\end{pmatrix} \\
F_{N-1,\text{MixCol}} &= S_{N-1,\text{MixCol}} + A_0 \cdot Z + A_0 \cdot \begin{pmatrix}
\epsilon & 0 & 0 & 0 \\
0 & 0 & 0 & 0 \\
0 & 0 & 0 & 0 \\
0 & 0 & 0 & 0
\end{pmatrix} \\
F_{N-1,\text{AddKey}} &= S_{N-1,\text{AddKey}} + A_0 \cdot Z + \begin{pmatrix}
2 \epsilon & 0 & 0 & 0 \\
\epsilon & 0 & 0 & 0 \\
\epsilon & 0 & 0 & 0 \\
3 \epsilon & 0 & 0 & 0
\end{pmatrix} \\
F_{N,\text{SubBytes}} &= S_{N,\text{SubBytes}} + \begin{pmatrix}
\epsilon'_0 & 0 & 0 & 0 \\
\epsilon'_1 & 0 & 0 & 0 \\
\epsilon'_2 & 0 & 0 & 0 \\
\epsilon'_3 & 0 & 0 & 0
\end{pmatrix} \\
F_{N,\text{ShiftRows}} &= S_{N,\text{ShiftRows}} + \begin{pmatrix}
\epsilon'_0 & 0 & 0 & 0 \\
0 & 0 & 0 & \epsilon'_1 \\
0 & 0 & \epsilon'_2 & 0 \\
0 & \epsilon'_3 & 0 & 0
\end{pmatrix} \\
F_{N,\text{AddKey}} &= S_{N,\text{AddKey}} + \begin{pmatrix}
\epsilon'_0 & 0 & 0 & 0 \\
0 & 0 & 0 & \epsilon'_1 \\
0 & 0 & \epsilon'_2 & 0 \\
0 & \epsilon'_3 & 0 & 0
\end{pmatrix}
\end{align*}

With $s(x)$ being the AES S-Box, we can find the following equations (on $x_0, x_1, x_2, x_3, \epsilon$)

\begin{equation}
\left \{ \begin{aligned}
s(x_0 + 2 \epsilon + 2 \zeta_{1} + 3 \zeta_{5} +   \zeta_{9} +   \zeta_{13}) & = \\
s(x_0 + 2 \zeta_{1} + 3 \zeta_{5} +   \zeta_{9} +   \zeta_{13}) + \epsilon'_0 & \\
s(x_1 +   \epsilon +   \zeta_{1} + 2 \zeta_{5} + 3 \zeta_{9} +   \zeta_{13}) & = \\
s(x_1 +   \zeta_{1} + 2 \zeta_{5} + 3 \zeta_{9} +   \zeta_{13}) + \epsilon'_1 & \\
s(x_2 +   \epsilon +   \zeta_{1} +   \zeta_{5} + 2 \zeta_{9} +   \zeta_{13}) & = \\
s(x_2 +   \zeta_{1} +   \zeta_{5} + 2 \zeta_{9} +   \zeta_{13}) + \epsilon'_2 & \\
s(x_3 + 3 \epsilon + 3 \zeta_{1} +   \zeta_{5} + 2 \zeta_{9} + 2 \zeta_{13}) & = \\
s(x_3 + 3 \zeta_{1} +   \zeta_{5} + 2 \zeta_{9} + 2 \zeta_{13}) + \epsilon'_3 & \\
\end{aligned}
\right .
\end{equation}

The trick here is that all the $\zeta$ terms are constant and with a change of variable, we get Dusart's original equations.

\begin{equation}
\left \{ \begin{aligned}
s(x'_0 + 2 \epsilon) &= s(x'_0) + \epsilon'_0 \\
s(x'_1 +   \epsilon) &= s(x'_1) + \epsilon'_1 \\
s(x'_2 +   \epsilon) &= s(x'_2) + \epsilon'_2 \\
s(x'_3 + 3 \epsilon) &= s(x'_3) + \epsilon'_3 \\
\end{aligned}
\right .
\end{equation}

The upshot is that any implementation of Dusart's attack as well as Piret's improvements (which is what we used) can be applied unmodified\footnote{In this attack, it is no longer true that invalid candidates will yield no solution. We ran into this issue on a small percentage of key-slots we attacked.} on ciphertexts with static faults. This means we can make use of a greater number of faulty ciphertexts without having to keep sampling and changing the input until we hit a lucky 1 byte fault. Since we do not know which faulty ciphertext has only static faults, we try every possible combination of $3 {M \choose 3}$ groupings with one candidate as the static-only faulty text. With this one weird trick, we were able to fully recover the remaining non-master key-slots.

\subsection{Targeting master key-slots}

Up until this point, we focused on non-master key-slots. Recall that master key-slots have an additional level of obfuscation: the engine does not directly reveal the ciphertext output. Instead, the engine writes the output ciphertext to another key-slot (which is not readable) to be used as a new key. Luckily, this security measure has already been cracked by David ``Davee'' Morgan~\cite{Davee} who found a hardware vulnerability in Bigmac. The last ciphertext is not cleared from the engine's internal state and the next invocation of the engine with an input of size $< 16$ bytes will ``borrow'' the remaining bytes from the ciphertext of the last successful operation. We can extract the ciphertext output of a master key operation with the following steps:

\begin{enumerate}
    \item Perform a faulty AES-256 decrypt using the master key-slot and any slave key-slot as the destination. Due to the vulnerability, a copy of the faulty ciphertext will remain in the AES engine's internal state.
    \item Using a known fixed key, AES-128 encrypt a buffer of 4 bytes of \texttt{00} to memory. Save the resulting ciphertext, $C_3$.
    \item With the same fixed key, AES-128 decrypt $C_3$, which restores the internal state.
    \item Repeat steps 2-3 with 8 bytes and 12 bytes of \texttt{00} to produce $C_2$ and $C_1$.
    \item Finally, use the slave key-slot to encrypt 16 bytes of \texttt{00} to produce $C_4$
\end{enumerate}

We can then do a $2^{32}$ brute-force on $C_1$ to find the first 4 bytes of the faulty ciphertext (since we used a fixed key, and we know that 12 bytes of the input are \texttt{00}). Then we can use the 4 known bytes and $C_2$ to find the next 4 bytes and repeat with $C_3$ to find the next 4 bytes. Finally we brute force the ``key'' used to produce $C_4$ with the first 12 bytes we found and a $2^{32}$ brute-force of the remaining 4 bytes. This gives a worst case of $4 * 2^{32} = 2^{34}$ AES operations to retrieve a single faulty ciphertext.

Using the \texttt{c5.18xlarge} instance on Amazon Web Services EC2 which provides 72 CPU cores~\cite{Amazon}, each faulty ciphertext retrieval takes an average of 15 seconds and the worst case of under a minute. After obtaining the faulty ciphertexts, we can perform the same DFA attack on the master slots as with other slots.

\subsection{Results}

With the phoenixAES library implementation of Piret's attack along with our brute-force enhancements, we able to use our round $N-2$ faults to obtain the round $N$ key and then use the round $N-3$ faults to obtain the round $N-1$ key.\footnote{With the round $N$ key, we can reverse a single round of AES for the correct ciphertext along with every faulty ciphertext. Then we just run the same DFA attack with the new sample set to get the round $N-1$ key.} Combining both gives us the full AES-256 key.

We were able to carry out the attack successfully on all $278$ key-slots we have access to (including all $30$ master key-slots). There are two key-slots, used for the device unique eMMC full-disk-encryption, that are locked out before we can gain code execution. In theory, it should be possible to perform the same attack by writing to the eMMC with our RPC (the FDE is done in hardware and is transparent to software) and dumping the result with an external flasher. However, we do not attempt this because of the extra overhead involved and the fact that the keys are device unique and therefore not useful to have.

\section{Conclusion}

We have demonstrated that AES DFA attacks do work well in practice, although some extra work was required. It is particularly attractive for devices like the PlayStation Vita where the software has been security-hardened. The entire cost of this attack was surprisingly low. ChipWhisperer Lite and CW501 differential probe costs about \$300. The custom boards and components for glitching and triggering costs less than \$10 total. 500 core-hours of partials busting on AWS EC2 costed us about \$10. Even including the extra equipment used during the development and debugging such as a \SI{100}{\mega\hertz} oscilloscope and extra Vita motherboards, the entire cost of the attack was easily under \$1000. We believe that all modern SoC should, if they do not already, defend against DFA attacks because these attacks are not just theoretical. The PlayStation Vita used hardware AES keys as a way of protecting the software, but because they did not also protect the hardware as well, all their defenses crumble with a precisely timed voltage spike.

\section*{Availability}

All our work are available as open source projects.

\begin{enumerate}
    \item The F00D RPC payload and ChipWhisperer scripts to run the RPC and collect faulty ciphertexts: \url{https://github.com/TeamMolecule/f00dsimpleserial/}
    \item AES fault analysis script for finding where the fault occurs given a known key: \url{https://github.com/TeamMolecule/f00dsimpleserial/tree/master/scripts/analysis}
    \item DFA attack script based on phoenixAES including the second order DFA enhancements: \url{https://github.com/TeamMolecule/f00dsimpleserial/tree/master/scripts/dfa_crack}
    \item Master key-slot ciphertext brute force with AES-NI support (thanks to ``Davee''): \url{https://github.com/TeamMolecule/f00d-partial-buster}
\end{enumerate}

\printbibliography

\onecolumn
\appendix
\section{Appendix}

\subsection{PCB Modifications}

\begin{figure}[h]
\centering
\includegraphics[width=12cm]{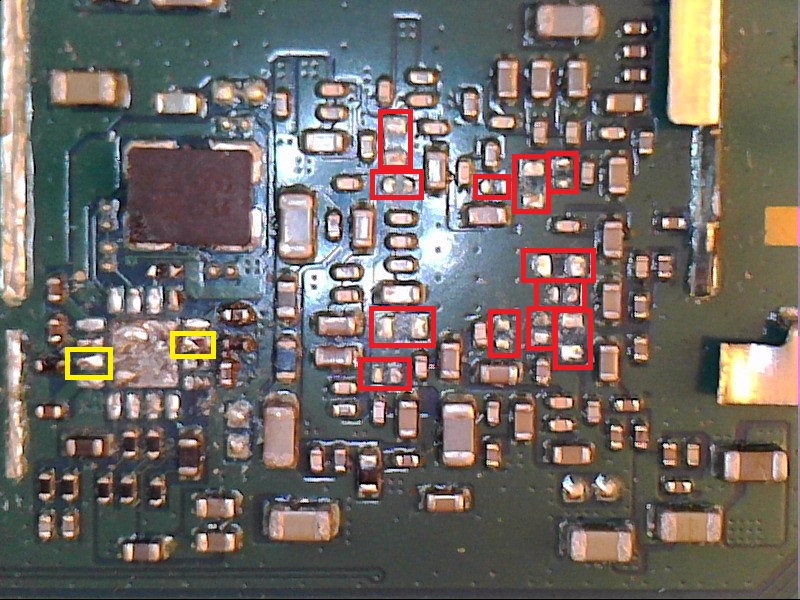}
\caption{The decoupling capacitors for the core 1.1V voltage domain removed are boxed in red. The two external clock input pads are boxed in yellow (the clock synthesizer chip is removed).}
\label{fig:decoups}
\end{figure}

\begin{figure}[h]
\centering
\includegraphics[width=12cm]{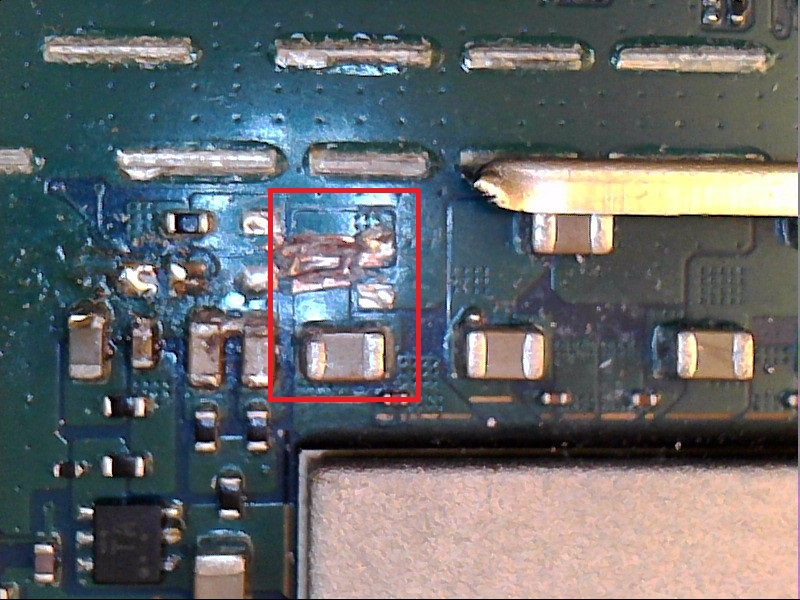}
\caption{The back of the board where the 1.1V supply trace is cut in order to isolate the regulator from the SoC.}
\label{fig:cuttrace}
\end{figure}

\begin{figure}[h]
\centering
\includegraphics[width=12cm]{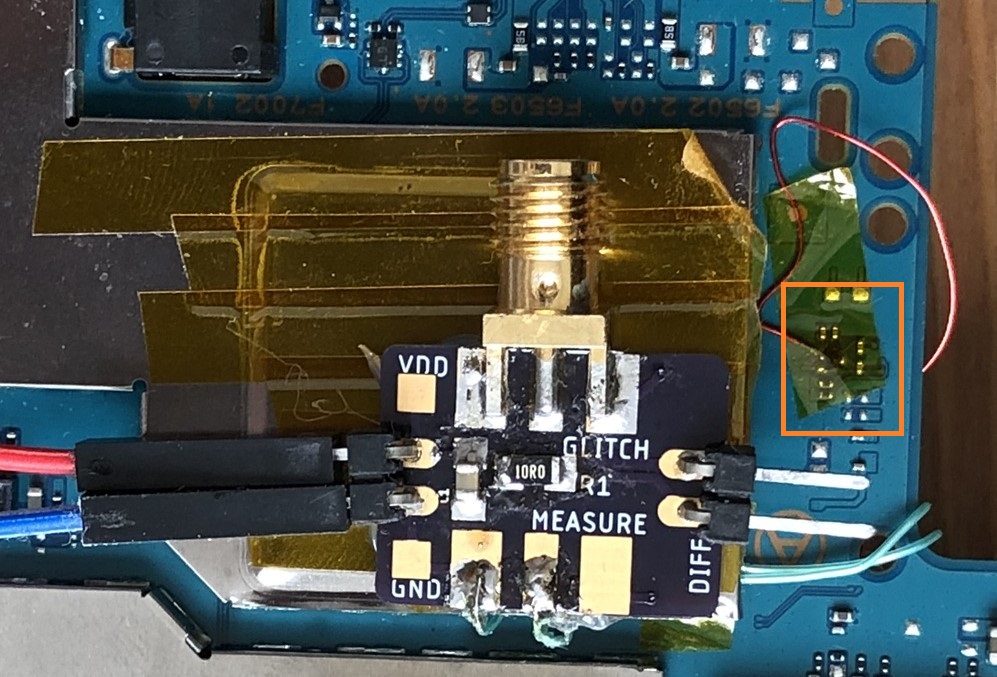}
\caption{The psvcw board glued to the under-side of the Vita motherboard. SMA connector goes to ChipWhisperer glitch module. The left pins go to an external 1.1V supply. The right pins go the differential probe. On the bottom of the board are two wires that are soldered to ground and the top portion of the cut trace. The shunt resistor is $\SI{10}{\ohm}$ and the bypass capacitor is $\SI{10}{\micro\farad}$. Boxed in orange is the GPIO output from the device used as a trigger.}
\label{fig:psvcw}
\end{figure}

\end{document}